\documentclass[ journal]{IEEEtran}
\usepackage{amssymb}
\usepackage{amsmath,amsfonts}
\usepackage[ruled,linesnumbered]{algorithm2e}
\usepackage{algorithmicx}
\usepackage{amsthm}
\usepackage{array}
\usepackage{textcomp}
\usepackage{stfloats}
\usepackage{url}
\usepackage{verbatim}
\usepackage{graphicx}
\usepackage{cite}
\usepackage{algpseudocode}
\usepackage{bm}
\usepackage{graphicx}
\usepackage{subfigure}
\usepackage{float}
\usepackage{booktabs}
\usepackage{color}
\usepackage{tikz}
\newcommand{\RNum}[1]{\uppercase\expandafter{\romannumeral #1\relax}}

\theoremstyle{definition}

\theoremstyle{remark}

\begin{document}

\title{{\it BS-1-to-N}: Diffusion-Based Environment-Aware Cross-BS Channel Knowledge Map Generation
 for Cell-Free Networks
}

 \author{{Zhuoyin Dai,  Di Wu,   Yong Zeng,~\IEEEmembership{Fellow, IEEE}, Xiaoli Xu,~\IEEEmembership{Member, IEEE},
 Xinyi Wang,~\IEEEmembership{Member, IEEE},
 and  Zesong Fei,~\IEEEmembership{Senior Member, IEEE}
 }  


 \thanks{
  Z. Dai, D. Wu,  Y. Zeng, and  X. Xu are with the National Mobile Communications Research Laboratory, 
  Southeast University, Nanjing 210096, China. Z. Dai, D. Wu,  and Y. Zeng  are also with the 
  Purple Mountain Laboratories, Nanjing 211111, 
  China (e-mail: \{zhuoyin\_dai, studywudi,  yong\_zeng,  xiaolixu\}@seu.edu.cn). 

  X. Wang and Z. Fei are with the School
  of Information and Electronics, Beijing Institute of Technology, Beijing 100081, China.
   (E-mail: bit\_wangxy@163.com, feizesong@bit.edu.cn)

  (\emph{Corresponding author: Yong Zeng.})

          }

} 

 \maketitle

\begin{abstract}
  Channel knowledge map (CKM) inference across base stations (BSs) is the key to achieving efficient environment-aware communications.
  This paper proposes an  environment-aware cross-BS CKM inference method called {\it BS-1-to-N} based on the generative  diffusion model. To this end, we first design the BS location embedding (BSLE) method tailored  for cross-BS CKM inference  to embed BS location information in the feature vector of CKM.
  Further,  we utilize the cross- and self-attention mechanism for the 
  proposed {\it BS-1-to-N} model to respectively  learn the relationships between source and target BSs, as well as that  among target BSs.
  Therefore,  given the locations of the source and target BSs, together with the source CKMs as control conditions, cross-BS CKM inference can be performed for an arbitrary number of source  and target BSs.
  Specifically, in architectures with massive distributed nodes like cell-free networks, traditional methods of sequentially traversing each BS for CKM construction are prohibitively costly.
  By contrast, the proposed {\it BS-1-to-N} model is able to achieve efficient CKM inference for a target BS at any potential location based on the  CKMs of  source BSs.
  This is achieved by exploiting the fact that within a given area, different BSs share the same wireless environment that leads to their respective CKMs.
  Therefore, similar to multi-view synthesis, CKMs of different BSs are representations of the same wireless environment from different BS locations. By mining the implicit correlation between CKM and BS location based on the wireless environment, the proposed {\it BS-1-to-N} method achieves efficient CKM inference across BSs.
  We provide extensive  comparisons of CKM inference between the proposed {\it BS-1-to-N} generative model versus benchmarking schemes, and provide one use case study to demonstrate its practical application for the  optimization of BS deployment.

\end{abstract}

\begin{IEEEkeywords}
  Channel knowledge map, environment-aware communication, generative diffusion model, 
  cross-BS CKM inference.
\end{IEEEkeywords}

\section{Introduction}
With the rapid development of wireless communication technology, future network 
architectures are evolving  toward  denser networks  with large-scale distributed nodes \cite{you2021towards}, such as cell-free networks \cite{ngo2017cell,dai2022rate} and 
ultra-dense networks (UDNs) \cite{kamel2016ultra,chen2016user}. 
For such  networks, achieving  environment-aware communication 
is expected to significantly improve the performance, as it can effectively utilize  
the macro diversity gains 
brought by distributed nodes with prior environment knowledge. 
Channel knowledge map (CKM) \cite{zeng2021toward} is an effective technology to achieve the above goals, 
as it learns 
the location-specific prior  channel knowledge
of various kinds, such as channel gain, multi-path delays,  angle of departure (AoD), and angle of arrival (AoA),  laying 
the foundation for environment-aware wireless communication and sensing \cite{Zeng2023ATO}.
Specifically, CKM can help  nodes such as base stations (BSs) and access points (APs) 
improve their awareness of the wireless environment, thereby efficiently performing 
typical tasks such as beam alignment \cite{wu2022environment}, resource allocation 
\cite{zeng2021toward}, non-line-of-sight (NLoS)  sensing  \cite{wu2025you}, and localization \cite{zhaole2025imnet}.

The key to achieving environment-aware communication lies in constructing the corresponding CKM 
for each BS, a task that becomes increasingly challenging for cell-free networks due to the high BS density.
Traditional CKM construction methods often focus on intra-BS  construction, i.e., constructing the CKM of a target BS based on multi-modal data associated with the same BS \cite{levie2021radiomap}.
To this end, it requires collecting channel knowledge data via either on-site measurements or   ray tracing computation  \cite{Hoydis2023sionna} through traversing all BSs, which is   costly and inefficient  for large-scale distributed nodes.
In particular, interpolation techniques such as expectation maximization (EM) \cite{li2022channel} and Gaussian process regression (GPR) \cite{dal2022modelfree} also require the traversal 
process, and their accuracy depends on the density of the sample data.
On the other hand, neural network-based methods, like RadioUNet \cite{levie2021radiomap}, 3D-RadioDiff \cite{zhaole20253d}, and 
CKMDiff \cite{fu2025ckmdiff}, usually focus on 
constructing intra-BS CKM based on physical environment maps or sparsely sampled data.
Meanwhile, the intra-BS CKM construction problems are usually modelled as linear inverse problems   that include 
CKM super-resolution \cite{zhang2023RME-GAN,wang2024deep}, inpainting, denoising, or even generation \cite{jin2025i2i} as special cases.
Overall, intra-BS CKM construction often treats the construction of  CKMs of different BSs as 
independent tasks, ignoring their interrelationships.

A notable feature of dense networks is that  BSs in an area at least partially or even  fully share the same physical environment. Note that the physical environment 
determines the wireless propagation characteristics, with channel knowledge being the direct representation.
Therefore, CKMs corresponding to different BSs can be essentially regarded as multiple perspectives 
of the same  wireless environment from different BS locations.
This characteristic of dense networks brings a new opportunity, i.e., cross-BS CKM 
inference, in which the CKM of the target BS at any potential new location can be inferred
based on CKMs of the source BSs, eliminating the need for repeated on-site  measurements or ray tracing computation  for each BS. 
There have been some preliminary efforts toward this direction. For example, 
channel-to-channel mapping was proposed for specific locations with deep neural networks \cite{chen2019learning, Alrabeiah2019deep}, but it has not exploited the  location information
and is not suitable for the construction of a complete CKM.
The structure of UNet-based cross-BS CKM inference \cite{dai2024generating}, including the number of inputs and outputs, is fixed, making it unsuitable for scenarios with flexible numbers of  source  and/or target BSs.
Furthermore, while diffusion models such as the latent diffusion model  (LDM) have been initially applied 
to intra-BS CKM construction \cite{fu2024generative}, they cannot be directly applied to cross-BS CKM inference, due to the lack of  a mechanism for 
characterizing the  relationship  among CKMs of different BSs.

To tackle the above challenges, this paper develops an efficient and flexible model, called {\it BS-1-to-N},  for
cross-BS CKM inference based on the generative  diffusion model.
The term ``{\it BS-1-to-N}'' is used to indicate that  the proposed model  enables cross-BS CKM inference for an arbitrary number of source BSs and target BSs, i.e., from 1 to $N$ source BSs and target BSs.
The core idea draws inspiration from multi-view synthesis \cite{Liu2023zero,Kong2024eschernet} in the image processing community, where different views of an 
object jointly characterize its features.
Similarly, CKMs of different BSs in the same area can be regarded as multi-view 
projections of the same wireless environment, each capturing and learning partial 
features of the wireless environment from the corresponding BS location.
This indicates that by exploiting  the intrinsic correlation between  CKMs of 
different BSs, the CKM of the target BS at any potential location in the area can be inferred based on those of the source BSs.
It is worth noting that this method does not require the physical environment 
map as   input, but only needs  CKMs and location information of the source BSs to infer 
the CKM of the target BS at any potential location.
In addition, the proposed cross-BS CKM inference method not only significantly 
reduces the CKM construction and updating cost, but also effectively guides 
applications such as the optimization of BS deployment.
The main contributions of this 
work are summarized as follows:
\begin{itemize}
\item
A generative cross-BS CKM inference framework called {\it BS-1-to-N} based on LDM  is proposed,
which formulates the cross-BS  CKM problem  as a conditional generative problem.
Without   relying on the physical environment map information, this 
framework can infer the  CKMs of  target BSs at any potential locations, significantly improving inference efficiency and scenario adaptability.

\item	
We develop  a BS location encoding (BSLE) mechanism that efficiently encodes BS location information and embeds it into CKM feature vectors during the diffusion process. 
BSLE achieves accurate correlation between location information and CKMs, laying a foundation for fusing local features of the wireless environment embodied in different CKMs.

\item	  
An inter-CKM attention mechanism is introduced, which learns  the
cross-attention between the source BSs and target BSs, and the self-attention between  the target BSs. 
This mechanism improves the accuracy  of CKM inference through information collaboration between multiple BSs and supports CKM inference from any number of source BSs and target BSs at the system level.

\item  The proposed {\it BS-1-to-N} model is trained using data from a certain number of source BSs. Extensive comparisons with benchmark methods, such as UNet-based inference, validate the effectiveness of generative cross-BS CKM inference. Furthermore, the  practical application of the proposed BS-1-to-N generative model is demonstrated through a typical case study on BS deployment optimization.

\end{itemize}

The remainder of this paper is structured as follows: Section II provides an overview 
of  the construction of CKMs and the fundamentals of diffusion models.
The problem formulation and the feasibility analysis of cross-BS CKM inference 
are given in Section III.
Section IV provides a detailed discussion of the architecture and mechanisms of {\it BS-1-to-N}.
Section V presents experiment results to validate the performance and demonstrate the  application of 
{\it BS-1-to-N} for BS deployment optimization. Finally, the   work is concluded in Section VI.

\emph{Notations:}  
Vectors and matrices are represented in  boldface  letters.
We use $\mathbb{R}^{N\times M}$ to denote the $N\times M$-dimensional real-valued matrix.
The general function is denoted by $f(\cdot )$ while the
probability function is denoted by $p(\cdot )$.
The dot-product between $\mathbf{a}$ and $\mathbf{b}$ is $\langle \mathbf{a}, \mathbf{b} \rangle$.
The transpose   of   $\mathbf{a}$ is denoted as $\mathbf{a}^{\mathrm{T}}$. 


\section{Preliminaries and Related Works}

\subsection{Channel Knowledge Map Construction}

Efficient  and high-quality  CKM  construction is the foundation for CKM-enabled 
environment-aware wireless communication and sensing.
The first issue that needs to be addressed for the construction of CKM is 
the acquisition of the initial channel knowledge when a new BS is to be deployed.
On-site  measurement and ray tracing computation are the two major methods
of channel knowledge acquisition.
On-site measurement is performed to obtain actual values of a wide variety of 
channel knowledge, such as path loss, time of arrival (TOA),  AOA,  AOD, 
etc.,  usually performing  measurements at discrete  locations.
In comparison, 
ray tracing computes the   wireless propagation characteristics  at different locations 
to obtain the corresponding channel knowledge.

The channel knowledge obtained from both on-site measurement and ray tracing computation  can be directly stored 
as location-specific data for CKMs.
However, both  on-site measurement and ray tracing computation
require significant time costs and computational resources, hindering their feasibility to construct high-resolution  CKMs.
One possible solution is to estimate or infer channel knowledge for those locations without any data from on-site measurement and ray tracing computation.
Interpolation with K-nearest neighbors is a typical technique for channel knowledge inference  in CKM, 
where the channel knowledge of the target locations is estimated by weighted summation of the data points of the $K$ nearest locations \cite{cover1967nearest}.
Further, the  EM algorithm is used to select statistical channel 
parameters and perform location-specific model-based channel knowledge interpolation \cite{li2022channel}.
Besides, to achieve model-free CKM estimation with high accuracy, a sampling-based  GPR   strategy is proposed for the joint estimation of grid-of-beams and path loss \cite{dal2022modelfree}.
Meanwhile, the  spatial  sampling density required to achieve CKM construction with the desired normalized mean square error (NMSE)
has been analyzed  based on stochastic geometry in \cite{xu2023much}.

Besides the aforementioned approaches that directly estimate the channel knowledge 
relying on on-site measurement and ray tracing computation,  neural network-based CKM generation has also 
attracted significant  attention.  These methods typically utilize the existing channel knowledge data
for neural network training based on different construction requirements,
and then generate corresponding CKMs in conjunction with physical or wireless 
environment characteristics. 
Note that CKM is not only UE location-specific but also BS-specific. 
Therefore,  neural network-based CKM generation can be categorized into
intra-BS generation and cross-BS generation, depending on whether the BSs associated with the input and output data are the same. 
As  a representative intra-BS generation approach, RadioUNet employs  both  physical environment map and  BS location as inputs to map the CKM of the BS based on the cascaded UNet network \cite{levie2021radiomap}.
To generate fine-resolution CKM with nonuniform sparse samplings, radio map estimation generative adversarial network (RME-GAN) focuses on the 
local features to  accommodate irregular samplings and outputs the CKM with high accuracy \cite{zhang2023RME-GAN}.
Further, the intra-BS construction is modeled as a  linear inverse problem in
CKMDiff \cite{fu2024generative},  using the sparsely observed channel knowledge data as the prompts for 
conditional generation of the CKM.
Problems such as intra-BS complementation \cite{wang2024radiodiff} and 
super-resolution   CKM  construction \cite{wang2024deep}  are also achieved by deep learning.
As for CKM construction across BSs,
Channel-to-channel mapping and inference are conducted for specific locations via
deep neural networks \cite{chen2019learning}.
Furthermore, a pioneering attempt in cross-BS CKM inference has been investigated in \cite{dai2024generating}, with the location and CKM information with different modalities directly weighted
and fed into the UNet with a fixed number of channels. 

\subsection{Latent Diffusion Model}

As a typical generative model, diffusion learns the distribution of the data through a denoising process, 
which is essentially the reverse  process of Markov chains \cite{ho2020denoising}.
Specifically, the denoising diffusion probability model (DDPM) divides the image generation  
into two processes, i.e., the forward diffusion process and the reverse  generation process. 
The forward diffusion process is essentially a Markov chain in which the data 
is transformed into Gaussian noise by continuously adding noise to it. 
In the reverse  generation process, the original data is recovered from Gaussian noise 
by training a neural network to predict and remove the noise added at each step.

Diffusion model has become a strong competitor to  GAN  in areas 
such as computer vision \cite{Dhariwal2021diffusion} due to its strong capabilities in image segmentation, repair,  inpainting,  decompression, and super-resolution
capabilities \cite{saharia2022palette}.
However, the promotion and application of traditional diffusion models also face bottlenecks.
On the one hand, the forward diffusion and reverse  generation processes of traditional diffusion models are typically operated directly on original high-dimensional images, which greatly limits the 
training efficiency and generation rate, and requires heavy hardware costs. On the other hand, traditional diffusion models usually implicitly fuse 
the condition injection with the noise prediction function through strategies 
such as classifier-guidance \cite{ho2022classifier}, which suffers from large parameter training, 
low applicability of multi-modal conditions, and low accuracy of condition control.

To address  the above problems, LDM has been proposed to achieve high-quality model training  with limited computational resources \cite{rombach2022high}.
Through introducing  pre-trained autoencoders for feature extraction  and alternatively training over the low-dimensional latent space, LDM is able to achieve a balance between the generation 
performance and the training complexity of the diffusion model.
Meanwhile,  LDM encodes multi-modal inputs such as text, image, and audio 
into the latent vector space through modal coding, acquiring a good 
multi-modal input capability. 
By further introducing the cross-attention mechanism into the diffusion model architecture,
LDM can learn the latent correlations among different modalities and 
achieve   generation  with the fusion control of multi-modal conditions.
Stable Diffusion is based on the framework of LDM and has become a breakthrough 
application of generative models. 
By integrating the contrastive language-image pre-training (CLIP) text encoder and 
focusing  on text-to-image generation tasks, Stable Diffusion represents a concrete implementation
of LDM and has promoted the widespread use of diffusion models.

\section{Problem Formulation for Cross-BS CKM Inference}

\begin{figure}[!t]
  \centering
    {\includegraphics[width=0.85\columnwidth]{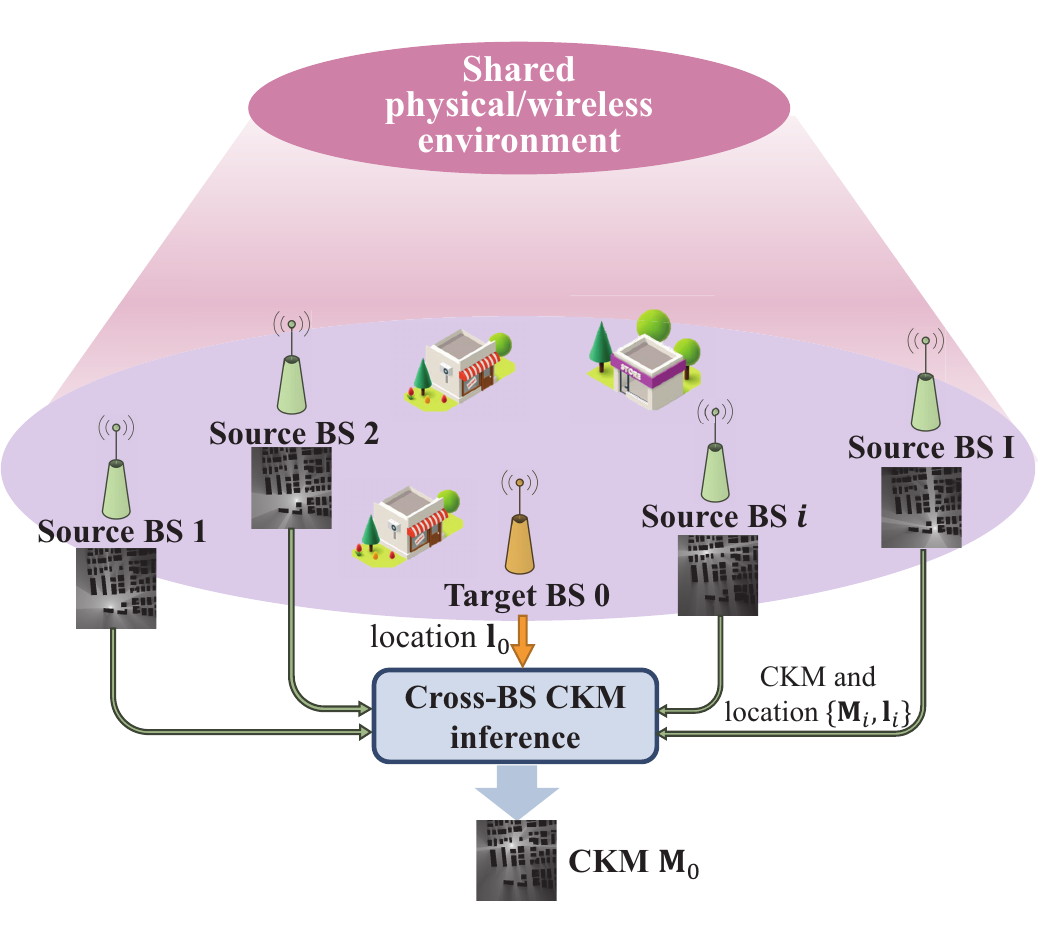}}%
  \caption{An illustration of cross-BS CKM inference problem. }
  \label{fig:mapping}
\end{figure}

We consider a system with distributed BSs, such as cell-free networks, as shown in Fig. 1.
We assume that all source BSs in the network have already learned their own  CKMs,  and new BSs need to be deployed during  the evolution of the cell-free network.
For a newly introduced target BS, inferring its CKM at different potential 
locations is significant not only for environment-aware  communication but also for applications such as
BS site selection, cell coverage prediction, and so on. For cross-BS CKM inference, this problem is formulated as 
inferring the CKM of the newly introduced target BS from the CKMs of other source BSs,
based on the environment awareness embedded in source  CKMs.

Specifically, the target area is divided into $L\times L$ grids, and 
each CKM $\mathbf{M}_{i}\in \mathbb{R}^{L\times L}$ stores a certain type of 
channel knowledge for the BS $i$ at location $\mathbf{l}_i$.
The cross-BS CKM inference problem  for the target BS 0 at location $\mathbf{l}_0$ is expressed as 
\begin{equation}\label{eq:fM0}
  f:\big(\{\mathbf{M}_{i},\mathbf{l}_{i} \}_{i=1}^{I}; \mathbf{l}_{0}\big) \rightarrow  \mathbf{M}_{0}.
\end{equation}

Although (\ref{eq:fM0}) resembles the typical problem corresponding to 
supervised learning \cite{dai2024generating}, the performance of traditional supervised learning is limited.
The  supervised learning model structure with scalability is difficult 
to design since the number of source BSs and target BSs may vary in practical scenarios.
Besides, there exist  correlations between different input elements 
(e.g., $\mathbf{M}_{i}$ and $\mathbf{l}_{i}$ of BS $i$) as well as different output 
elements (e.g., $\mathbf{M}'_{0}$ and $\mathbf{M}''_{0}$ of different target BSs), which are difficult to learn via traditional supervised learning  methods.

\subsection{Problem Analysis}

Before further   reformulation of the problem of cross-BS CKM inference, 
a feasibility analysis is first performed.
The essence of cross-BS CKM inference is to explore  the spatial correlation of CKMs of 
BSs at different  locations.
This correlation stems from the fact that the CKMs of different BSs share the same area 
and therefore rely on the same overall wireless environment for propagation.
Specifically, let $ c_{i,u}$ denote the channel knowledge between BS $i$ and a user 
located at $\mathbf{l}^{\mathrm{u}}$, which can be represented as \cite{Zeng2023ATO}
\begin{equation}\label{eq:ct}
  c_{i,u}=f_c(E,\mathbf{l}^{\mathrm{u}};\mathbf{l}_i), \forall u, i,
\end{equation}
where $E$ denotes the abstractive wireless environment of the whole area.
The environment $E$ is considered as a quasi-static environment for ease of illustration.
(\ref{eq:ct})  shows the theoretical possibility of inferring the 
corresponding channel knowledge from the known BS location $\mathbf{l}_i$, 
user location $\mathbf{l}^{\mathrm{u}}$, and wireless environment $E$.
However, as revealed in \cite{Zeng2023ATO},  the abstractive wireless environment encompasses 
the physical environment and radio propagation features, which is difficult to mathematically model. Furthermore,  it is mathematically difficult to find
a universal function $f_c(\cdot,\cdot;\cdot )$  
that accurately reflects the mapping relationship between the
location information  and the channel knowledge based on 
the wireless environment. 

Hence, in this paper, instead of explicitly modeling the wireless environment $E$ or the function $f_c$, we execute
the cross-BS CKM inference  problem  through mining the correlation of CKMs between different BSs
sharing the same wireless environment.
Specifically, each element with row $j$ and column $k$ of the   CKM $\mathbf{M}_{i}$ is the channel knowledge
corresponding to the user location $\mathbf{l}^{\mathrm{u}}_{j,k}$ as 
\begin{equation}
  [\mathbf{M}_{i}]_{j,k} =f_c(E,\mathbf{l}^{\mathrm{u}}_{j,k};\mathbf{l}_i).
\end{equation}

Therefore, each  $\mathbf{M}_{i} $ stores  location-channel knowledge pairs about the 
BS $i$, where each pair is a sampled realization of the channel knowledge mapping in (\ref{eq:ct}).
With   sufficient BSs at location $\{ \mathbf{l}_{i} \}_{i=1}^{I}$ and 
the corresponding CKMs $\{\mathbf{M}_{i}  \}_{i=1}^{I}$, it is theoretically possible to 
reconstruct  the wireless environment $E$,
which is basically the inverse problem of (\ref{eq:ct}) and expressed as 

\begin{equation}\label{eq:hatE}
  \hat{E}=f_e(\{\mathbf{M}_{i},\mathbf{l}_{i} \}_{i=1}^{I}).
\end{equation}

\begin{figure}[!t]
  \centering
    {\includegraphics[width=1\columnwidth]{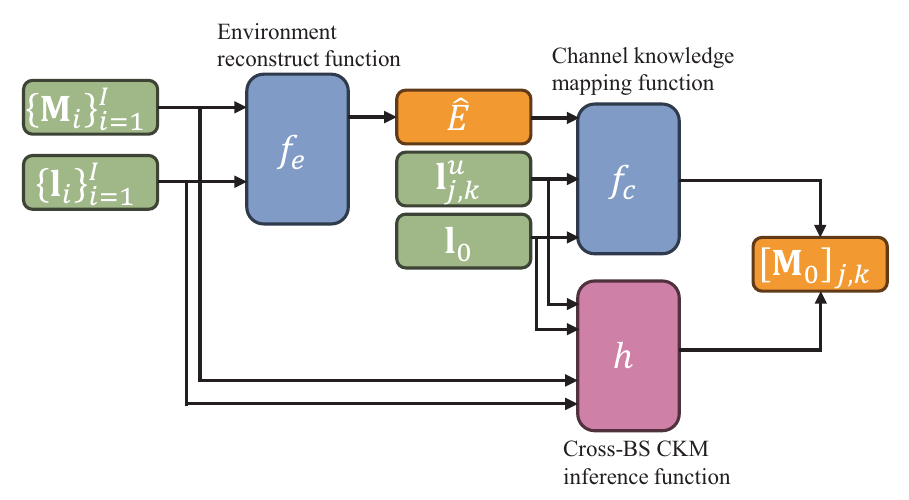}}%
  \caption{An Illustration of the feasibility of cross-BS CKM inference.}
  \label{fig:feasibility}
\end{figure}

Further, consider a  BS 0 to be newly deployed, which needs to
substitute into (\ref{eq:ct}) with $\hat{E}$ in (\ref{eq:hatE}), its location-specific 
channel knowledge in CKM $\mathbf{M}_{0}$ can be represented as
\begin{equation}
  \begin{aligned}\label{eq:Mijk}
    {[\mathbf{M}_{0}]}_{j,k} &=f_c(\hat{E},\mathbf{l}^{\mathrm{u}}_{j,k};\mathbf{l}_0) \\
     &=f_c\big(f_e(\{\mathbf{M}_{i},\mathbf{l}_{i} \}_{i=1}^{I}),\mathbf{l}^{\mathrm{u}}_{j,k};\mathbf{l}_0\big)\\
  & =h(\{\mathbf{M}_{i},\mathbf{l}_{i} \}_{i=1}^{I};\mathbf{l}^{\mathrm{u}}_{j,k},\mathbf{l}_0),
    \end{aligned}
\end{equation}
where $h(\cdot;\cdot,\cdot)$ is formed by the concatenation  of $f_c$ and $f_e$, showing the 
location-specific form of the CKM inference across BSs in (\ref{eq:fM0}).
The transformation process from (\ref{eq:ct}) to (\ref{eq:Mijk}) is shown in Fig. \ref{fig:feasibility}.
From Fig. \ref{fig:feasibility}, the mapping in (\ref{eq:Mijk}) avoids the explicit  reconstruction of the wireless environment,
demonstrating the feasibility of direct cross-BS  inference based on correlation between CKMs.

\begin{figure}[!t]
  \centering
    {\includegraphics[width=1\columnwidth]{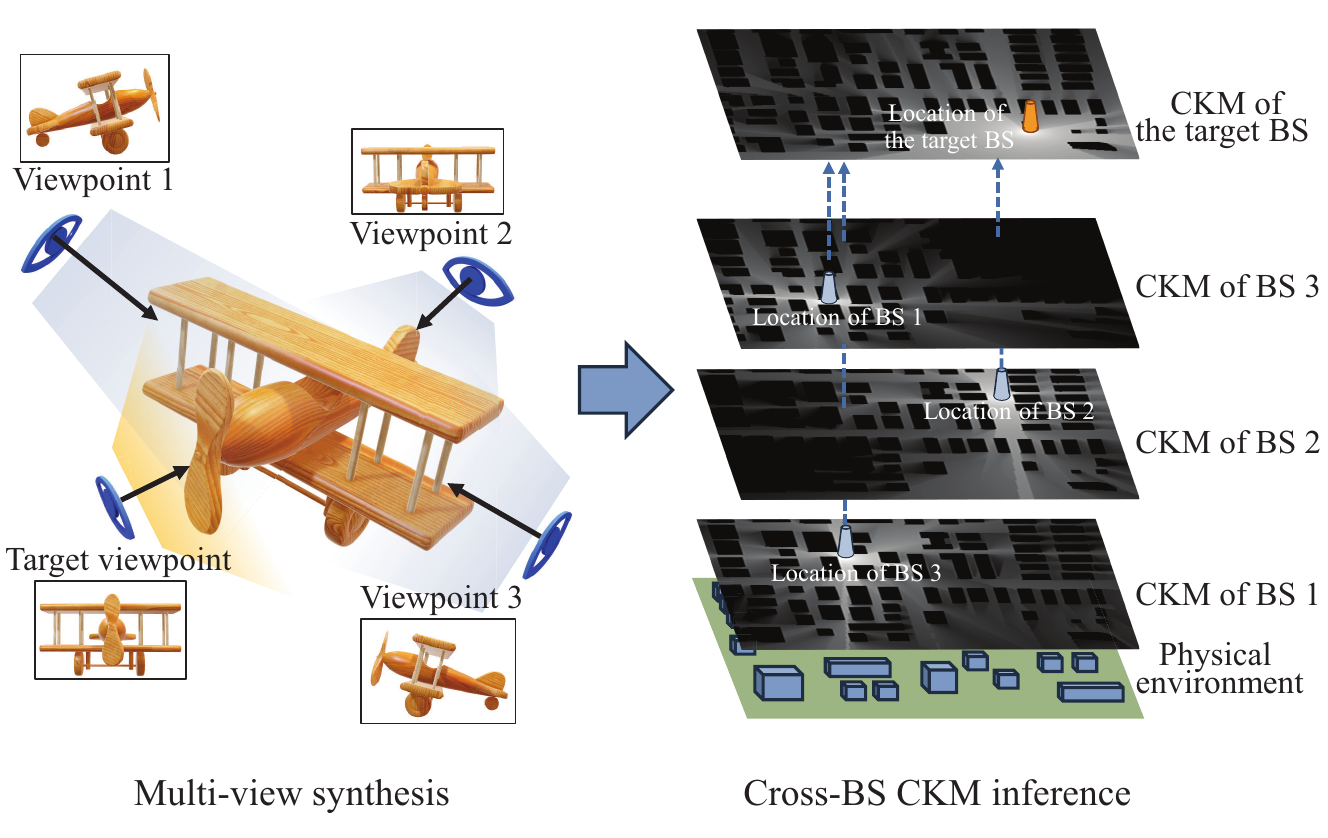}}%
  \caption{Analog  of multi-view synthesis and cross-BS CKM inference.}
  \label{fig:comparison}
\end{figure}

Although the abstractive features of the wireless environment make it difficult to 
derive and characterize correlations between CKMs of different BSs, scalable view 
synthesis in the image processing community  brings a novel idea.
For a physical  object,  observations  from  different viewpoints share 
some  properties about the object itself. 
As shown in Fig. \ref{fig:comparison}, the properties of the 
observed object will be better sensed as more views from different viewpoints are provided.
By encoding existing source viewpoints and the corresponding views as training prompts,
it becomes possible to 
generate views for arbitrary target viewpoints with deep neural networks \cite{Kong2024eschernet}.

Notice that CKM is a concrete and numerical manifestation of the wireless environment.
Therefore, for the cross-BS CKM inference problem, by treating  the wireless environment as the analogue of the object to be observed in multi-view synthesis, we can accordingly regard 
the CKMs of BS at different locations as projections or characterizations of the wireless
environment from different viewpoints, where the differences in viewpoints are reflected in
BS locations.
The CKM of each BS can provide partial awareness of the same wireless environment 
embedded in the channel knowledge from its location/viewpoint,
while jointly mining the channel knowledge  of multiple BSs can improve the awareness of the overall wireless environment.
Therefore, cross-BS CKM inference can be further formulated
as the following conditional generative modelling problem  for obtaining the distribution 
\begin{equation} \label{eq:problemconditional}
  \mathbf{M}_{0} \sim p \big(\mathbf{M}_{0} |    
  \{\mathbf{M}_{i},\mathbf{l}_{i} \}_{i=1}^{I}; \mathbf{l}_{0}   \big),
\end{equation}
where the number $I$ of the source CKMs can be arbitrary in both the model training and the inference phases.

\subsection{BS Location/Viewpoint Encoding}
CKM will be transformed into the intermediate representation through specific image encoders or in UNet layers.
To explore the location-specific intrinsic correlation between CKMs, the BSLE needs to be designed
to  embed BS locations accurately into the CKM tokens of  target BSs  and
source BSs. The differences between the encoding in CKM and  other domains are shown in
Table 1. 
Since CKM presents channel knowledge for a specific BS in a bounded area, it is 
different from the unbounded camera viewpoints in scalable view synthesis. 
However, the intrinsic correlation between the channel knowledge of different BSs 
still depends on their relative BS locations, which are determined by the 
relative locations between the physical environments in which the BSs are located.
Therefore, we design the following BSLE based on the polar coordinate system, which embeds global BS location information into  tokens to capture the relative BS location 
within the dot-product of attention mechanisms.

\begin{table}[H]
  \caption{Differences between Encodings in Different Domains }
  \label{table1p}
  \centering
  \resizebox{0.95\columnwidth}{!}{
  \begin{tabular}{c|cccc}
  \hline
      & Object &Position& Structure&Boundary  \\
  \hline
  Language& word & discrete &linear&bounded \\
  3D vision \cite{Kong2024eschernet}& camera & continuous & cyclic & unbounded  \\
  CKM& BS &continuous&cyclic&bounded  \\
  \hline
  \end{tabular}}
  \end{table}

\begin{figure}[!t]
  \centering
    {\includegraphics[width=0.6\columnwidth]{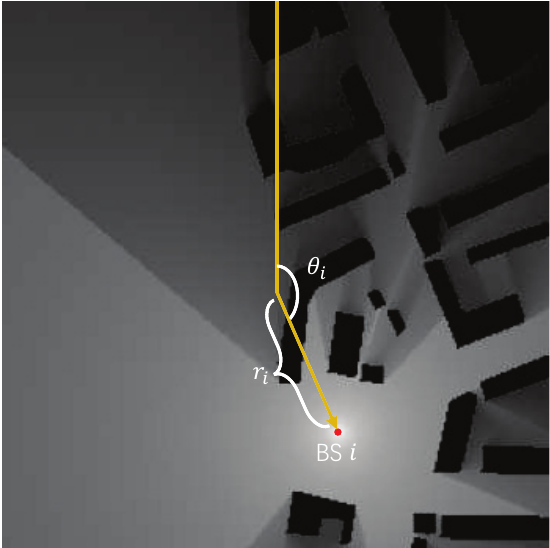}}%
  \caption{Illustration of the BS location encoding (BSLE).}
  \label{fig:BSLE}
\end{figure}

As shown in Fig. \ref{fig:BSLE}, in the considered polar coordinate system, the center of the  CKM area
is designated as the pole, and the polar axis is defined as the ray extending directly upward from this pole. Therefore, the location of BS $i$ is denoted as $\mathbf{l}_i=\{\theta_i,r_i \}
$, where $\theta_i\in [0,2\pi)$ is the polar angle, 
$r_i \in [0,r_{{\rm max}}]$ is the polar radius with $r_{{\rm max}}=\frac{\sqrt{2}}{2}L$.
In this design, the polar radius $r_i$ of the BS position $\mathbf{l}_i$ is bounded according 
to the boundary of the CKM area, while the polar angle $\theta_i$ is cyclic.

Subsequently, we investigate the encoding of  BS locations in combination with their corresponding CKM intermediate representations,
thereby enabling the joint feeding of multi-modal CKM and location information   into the model as tokens for training and inference.
Denote by $\mathbf{v}_i \in \mathbb{R}^{d\times 1}$ the
CKM token feature of BS $i$   and by $g(\cdot,\cdot)\in \mathbb{R}^{d\times 1}$ the location encoding
function that embeds the location 
into the  CKM token.
To ensure that the dot-product $ \langle\cdot, \cdot \rangle$  in  attention mechanisms is solely dependent   on the 
relative location of BSs, the following relation should be satisfied
\begin{equation} \label{eq:gtheta}
   \langle g(\mathbf{v}_i,\theta_i),  g(\mathbf{v}_j,\theta_j) \rangle
   = \langle g(\mathbf{v}_i,\theta_i-\theta_j),  g(\mathbf{v}_j,0) \rangle,
\end{equation}
\begin{equation}\label{eq:gr}
  \langle g(\mathbf{v}_i,r_i),  g(\mathbf{v}_j,r_j) \rangle
  = \langle g(\mathbf{v}_i,r_i-r_j),  g(\mathbf{v}_j,0) \rangle.
\end{equation}  

To ensure (\ref{eq:gtheta}) and (\ref{eq:gr}), the rotary position embedding strategy 
in RoFormer \cite{SU2024roformer} is adopted.
Different from other absolute or relative location encoding schemes that  often  rely on adding
the location encoding directly to the token, RoFormer embeds the absolute location information
in the rotation matrix, allowing the attention mechanism to exhibit an explicit relative location dependence.
While performing a rotary embedding of the BS location into the CKM token, it should be noted that
although the forms of (\ref{eq:gtheta}) and (\ref{eq:gr}) are the same, the pole angle $\theta_i$
and the pole axis $r_i$ have different physical meanings and ranges.
In addition, it should be considered that $\theta_i$ is cyclic while $r_i$ is bounded.
As a result, to encode (\ref{eq:gtheta}) and (\ref{eq:gr}) in a uniform form, the polar radius $r_i$
is first projected to the  angular domain as
\begin{equation}\label{eq:frri}
  f_{r}(r_i)=\frac{r_i}{r_{{\rm max}}} \pi \in[0,\pi].
\end{equation}

After the projection of $r_i$, we introduce the rotary position embedding strategy 
starting from the simple 2-dimensional case.

\subsubsection{2-dimensional rotary embedding}
Assume that the CKM token $\mathbf{v}_i$ has 2 dimensions with $\mathbf{v}_i=[v_{i1},v_{i2}]^T$.
Considering $\mathbf{v}_i$ as a vector in a 2-dimensional coordinate system, the 
rotation matrix corresponding to an 
$\alpha_i$-degree  rotation is
\begin{equation}
  \mathbf{R}_{\alpha_i}
  =\begin{bmatrix}
    \cos \alpha_i &-\sin \alpha_i \\
    \sin \alpha_i &\cos \alpha_i \\ 
    \end{bmatrix},
\end{equation}
and the location encoding function $g(\cdot,\cdot)$
that embeds the angle information into the CKM token can be  expressed as 
\begin{equation}
  g(\mathbf{v}_i,\alpha_i)=\mathbf{R}_{\alpha_i}\mathbf{v}_i.
\end{equation}

Based on the property of rotation matrix, the dot-product of $\mathbf{v}_i$ with $\alpha_i$-degree
and $\mathbf{v}_j$ with $\alpha_j$-degree
in attention mechanisms can be expressed as 
\begin{equation}
  \begin{aligned}
    \langle g(\mathbf{v}_i,\alpha_i),  g(\mathbf{v}_j,\alpha_j) \rangle
    &= (\mathbf{R}_{\alpha_i}\mathbf{v}_i)^{\mathrm{T}} \mathbf{R}_{\alpha_j}\mathbf{v}_j\\
     &=\mathbf{v}_i^{\mathrm{T}} \mathbf{R}_{\alpha_j-\alpha_i}\mathbf{v}_j \\
    &= \langle g(\mathbf{v}_i,\alpha_i-\alpha_j),  g(\mathbf{v}_j,0) \rangle,
  \end{aligned}
\end{equation}
which depends only on the relative angle $\alpha_i-\alpha_j$.

\subsubsection{ Rotary embedding of the BS location}
We denote the intermediate representation as a token $\mathbf{v}_i\in \mathbb{R}^{d\times 1}$ with $d$ 
dimensions.
Notice that the   BS location is characterized by $\mathbf{l}_i=\{\theta_i,r_i \}
$ in 2 dimensions according to the 
polar coordinate system.  Without loss of generality, 
it is assumed that $d$ is divisible by $2|\mathbf{l}_i|=4$.
Therefore, based on the rotation properties in the 2-dimensional
case, the following rotation matrix for BS $i$ with $\mathbf{l}_i$ can be obtained as 
\begin{equation}\label{eq:Ri}
  \mathbf{R}_i
  =\begin{bmatrix}
    \mathbf{R}(\mathbf{l}_i)  &\cdots &0\\
     \vdots  &\ddots   &\vdots \\ 
     0 &\cdots  &\mathbf{R}(\mathbf{l}_i) \\ 
    \end{bmatrix},
\end{equation}
where
\begin{equation}\label{eq:RIi}
  \mathbf{R}(\mathbf{l}_i) =\begin{bmatrix}
    \cos \theta_i & -\sin \theta_i &0  &0 &\\
    \sin \theta_i &   \cos \theta_i&  0& 0 &\\
      0& 0& \cos (f_{r}(r_i)) & -\sin(f_{r}(r_i))&\\
     0 & 0 &\sin (f_{r}(r_i)) & \cos (f_{r}(r_i))&\\
    \end{bmatrix}
\end{equation} 
consists of  $\theta_i $ and the projection $f_{r}(r_i)$ of  $r_i$ in the angular domain.

Therefore, based on (\ref{eq:Ri}) and (\ref{eq:RIi}), the  location encoding function 
is 
\begin{equation}\label{eq:gvi}
  g(\mathbf{v}_i,\mathbf{l}_i)=\mathbf{R}_i \mathbf{v}_i,
\end{equation}
where the attention between query and key can be calculated merely based on the 
relative location between BSs.
The entire process of BSLE is presented in Algorithm \ref{algorithm:1}.
\begin{algorithm}
  \caption{ BSLE Process of   BSs}\label{algorithm:1}
  \KwIn{the  CKM and location set $\{\mathbf{M}_{i},\mathbf{l}_{i} \}$}
  \KwOut{the   query matrix    $\tilde{\mathbf{Q}}$ in (\ref{eq:Qtilde})}
  Transform $\{\mathbf{M}_{i}\}$ into the intermediate representation $\mathbf{X}$ with the image
    encoder/UNet layer\;
    Construct the query matrix  $\mathbf{Q} $ with $\mathbf{X}$ based on the attention mechanism as discussed in Section IV\;
  \For{ \rm{each } BS $i$  }
  {Express the locations  in dataset as polar domain: $\mathbf{l}_i=\{\theta_i,r_i \}$\;
    Project $r_i$ into angle domain with (\ref{eq:frri}) to get $f_{r}(r_i)$\;
    Construct the rotation matrix $\mathbf{R}_i$ with (\ref{eq:Ri})\;
      Embed the corresponding BS location information with  (\ref{eq:gvi}) to get
      $\tilde{\mathbf{q}}_i$ in (\ref{eq:qttilde})
    }
  \end{algorithm}





\section{Diffusion-based CKM Inference across BSs}
In this section, details of how to realize CKM inference across BSs based on the 
diffusion model are presented.  This model is named {\it BS-1-to-N}, which means to learn the 
wireless environment characteristics from the CKMs of different numbers of BSs 
and perform CKM inference across any 
number of BSs for any number of target BSs. The whole network architecture is first presented. Subsequently,  the fundamentals of 
DDPM and the LDM methodology are  elaborated.
At last, the attention mechanism applied between different BSs and corresponding CKMs
is also presented.

\begin{figure*}[!t]
  \centering
    {\includegraphics[width=0.95\textwidth]{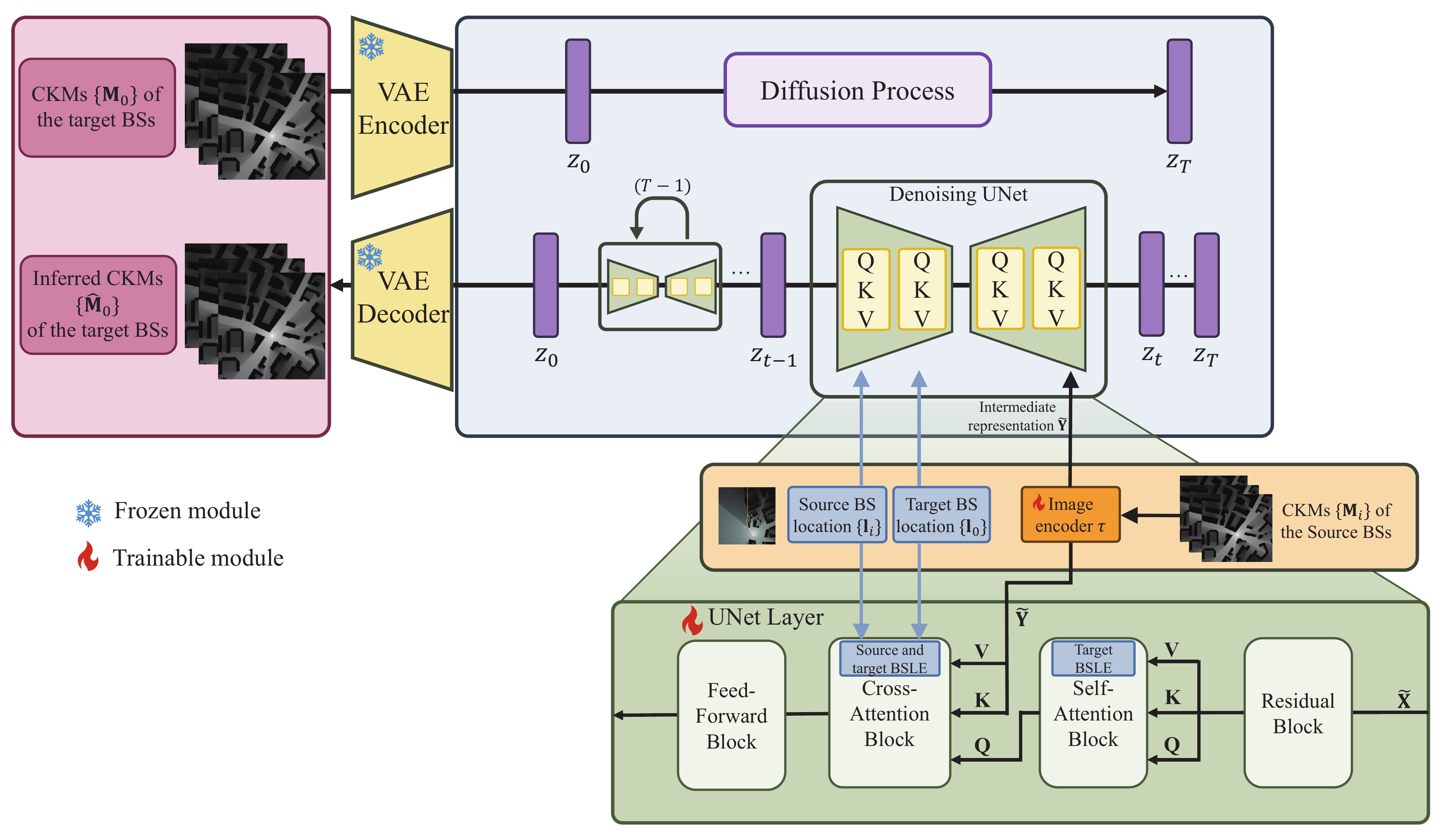}}%
  \caption{Architecture of the proposed {\it BS-1-to-N} generative diffusion model for cross-BS CKM inference.}
  \label{fig:architecture}
\end{figure*}

\subsection{Architecture Design}
The architecture of {\it BS-1-to-N} is built  on the basis of classic 2D diffusion models
and thus inherits its strong network architecture prior.
Further, similar to encoding the viewpoint pose for each view in multi-view synthesis, 
{\it BS-1-to-N} encodes the corresponding BS location for each CKM, thereby realizing any 
number of CKM inferences across any number of BSs. 
The basic architecture of {\it BS-1-to-N} is shown in Fig. \ref{fig:architecture},
where the variational autoencoder (VAE)  abstracts the CKM from  pixel space to latent space and the VAE decoder recovers
the latent space representation of the CKM to pixel space. 
In latent space, the diffusion module performs stepwise noise addition to the 
latent space input, while the UNet-based denoising module performs stepwise 
denoising generation.
The condition control module encodes the locations of  source BSs as well as their CKMs 
as prompts, and acts on the inference generation process of the CKM  of the target BS
through the attention mechanism. 
Note that the CKM $\mathbf{M}_0$ of the target BS is input during the training phase. In the inference phase, cross-BS inference starts directly from $\mathbf{z}_T$, without  the need to input $\mathbf{M}_0$.



DDPM is a classic implementation of diffusion models in the image field \cite{ho2020denoising}, 
whose core idea is to model the data distribution through the processes of forward 
noise addition and  reverse denoising, thus achieving high-quality image generation.
Taking a specific CKM $\mathbf{M}$ as an example, the following Markov chain is formed by gradually adding 
Gaussian noise 
to $\mathbf{M}$ with a variance sequence of $\{\beta_t\}$  during the forward process as 
\begin{equation}
  \mathbf{M}_t\sim q( \mathbf{M}_t| \mathbf{M}_{t-1})=\mathcal{N}(\mathbf{M}_{t};
  \sqrt{1-\beta_{t}}\mathbf{M}_{t-1},\beta_{t}\mathbf{I}).
\end{equation}
Based on the properties of joint probability density and Markov chain, 
the whole forward process can be expressed as the following posteriori estimation 
\begin{equation}
  q(\mathbf{M}_{1:T}|\mathbf{M}_{0})=\Pi_{t=1}^{T}q(\mathbf{M}_t| \mathbf{M}_{t-1}).
\end{equation}
Furthermore, according to the reparameterization trick, 
the closed-form sampling of the CKM $\mathbf{M}_t$ with noise at any timestep $t$ can 
be obtained as
\begin{equation}\label{eq:MtfromM0}
  \mathbf{M}_t \sim q( \mathbf{M}_t| \mathbf{M}_{0})=\mathcal{N}(\mathbf{M}_{t};
  \sqrt{\bar{\alpha_t}} \mathbf{M}_{0},(1-\bar{\alpha_t})\mathbf{I}),
\end{equation}
with $\alpha_t=1-\beta_{t}$ and $\bar{\alpha_t}=\Pi_{r=1}^{t}\alpha_r$.

In the reverse  denoising process, the final output of noise-free $ \mathbf{M}_{0}$ is 
obtained by gradually eliminating noise from $\mathbf{M}_t$.
The reverse denoising process can be modeled as  
\begin{equation}
  q(\mathbf{M}_{t-1}|\mathbf{M}_t)=\mathcal{N}(\mathbf{M}_{t-1};\mu_t(\mathbf{M}_t,
   \mathbf{M}_{0}),\sigma_t^2 \mathbf{I}),
\end{equation}
where according to Bayes theorem and (\ref{eq:MtfromM0}), 
$\mu_t(\mathbf{M}_t,\mathbf{M}_{0})=\mu_t(\mathbf{M}_t)=\frac{1}{\sqrt{\alpha_t}}\mathbf{M}_t
-\frac{1-\alpha_t}{\sqrt{1-\bar{\alpha_t}}\sqrt{\alpha_t}}\bm{\varepsilon}_t $
and $\sigma_t^2=\frac{(1-\alpha_t)(1-\bar{\alpha}_{t-1})}{1-\bar{\alpha_t}}$.

Considering that  $q(\mathbf{M}_{t-1}|\mathbf{M}_t)$ is difficult to derive,
the purpose of the reverse process is to train the corresponding neural network 
$p_{\vartheta}(\mathbf{M}_{t-1}|\mathbf{M}_t)$ to approximate $q(\mathbf{M}_{t-1}|\mathbf{M}_t)$.
Therefore, the loss function is defined as the negative likelihood function 
$-\log (p_{\vartheta}(\mathbf{M}_0))$.
Further, the loss function, which is difficult to optimize, 
can be approximated  by optimizing its variational lower bound.
After expanding the variational lower bound according to the total probability formula 
and neglecting the constant term, the new optimization objective is  obtained 
as the following  Kullback-Leibler (KL) divergence
\begin{equation}
  \arg \min_{\vartheta}D_{{\rm KL}}(q(\mathbf{M}_{t-1}|\mathbf{M}_t,\mathbf{M}_0)\|
  p_{\vartheta}(\mathbf{M}_{t-1}|\mathbf{M}_t) ).
\end{equation}

Note that the variance $\sigma_t^2$ is known. Hence, training the neural network to 
approximate the denoising process $q$ of the CKM is indeed approximating the 
mean $\mu_t$ of $q$, and at a more fundamental level, approximating the noise $\bm{\varepsilon}_t$ 
in $q$ by $\bm{\varepsilon}_{\vartheta}(\mathbf{M}_t,t)$. By comparing the mean square error (MSE) between the target mean and the 
approximate mean and neglecting the weighting constant coefficient, 
the simplified stepwise denoising loss can be obtained as
\begin{equation}
  L(\vartheta)=\mathbb{E}_{t,\mathbf{M}_0,\bm{\varepsilon}_t}\big[  \| \bm{\varepsilon}_t-
  \bm{\varepsilon}_{\vartheta}(\mathbf{M}_t,t)   \|^2  \big].
\end{equation}

\subsection{LDM in Cross-BS CKM Inference}

Although it is feasible to perform forward noise addition and reverse denoising
in the original pixel space of   CKMs, the high-dimensional pixel space imposes an unaffordable computational overhead.
Therefore, the diffusion process can be transferred from pixel space to latent space 
by introducing pre-trained perceptual compression models.
For example, on the basis of the encoder module $\varepsilon_{{\rm VAE}}$ in the VAE provided in Stable Diffusion,
the CKM $\mathbf{M}_0$ can be projected  to the latent space as 
\begin{equation}
  \mathbf{z}_{0}=\epsilon_{{\rm VAE}}(\mathbf{M}_0).
\end{equation}

In the process of compression, high-frequency but low-informative data in the pixel space 
are abstracted away, while low-frequency and high-informative data are given more 
attention.
Diffusion training in the low-dimensional latent space is more 
efficient due to  the decrease in dimensions.
Therefore, the compressed latent space is more suitable for likelihood-based 
generative tasks than the original pixel space.
In general, the derivation of the forward and reverse processes mentioned in the 
previous subsection still holds, while the object of operation is changed from $\mathbf{M}_t$
to $\mathbf{z}_{t}$. The corresponding denoising loss with timestep $t$  is rewritten as 
\begin{equation}\label{eq:loss}
  L(\vartheta)=\mathbb{E}_{t,\epsilon_{{\rm VAE}}(\mathbf{M}_0),\bm{\varepsilon}_t}\big[  \| \bm{\varepsilon}_t-
  \bm{\varepsilon}_{\vartheta}(\mathbf{z}_t,t)   \|^2  \big].
\end{equation}

The sample $\mathbf{z}_0$ is obtained through the reverse denoising process in the latent space. It can be further transferred back to the inferred CKM $\hat{\mathbf{M}}_0$ in the pixel 
space again via the decoding module $\mathcal{D}_{{\rm VAE}}$ in the VAE as 
\begin{equation}
  \hat{\mathbf{M}}_{0}=\mathcal{D}_{{\rm VAE}}(\mathbf{z}_0).
\end{equation}

\subsection{Attention Mechanism across BSs}

Stable Diffusion was originally designed for image generation guided by text conditions
and is difficult to directly apply to CKM inference across BSs. 
To realize the interactions between CKMs across BSs, the attention  computation
for multi-view synthesis in Eschernet \cite{Kong2024eschernet} is adapted to the attention  mechanism 
between CKMs corresponding to different BSs in {\it BS-1-to-N}.
Specifically, the self-attention module in Stable Diffusion, which is originally used for 
interrelationships within the same generative object, 
is  adjusted to learn the interactions between CKMs of different target BSs, 
thus maintaining consistency between CKM inference.
Meanwhile, the cross-attention between target and reference views in Eschernet 
is migrated to {\it BS-1-to-N} to ensure the consistency between the CKMs of the 
source BSs and the inferred CKMs of target BSs.
Details of the aforementioned adjustment are presented as follows:

\subsubsection{Self-Attention within Target CKMs}
The self-attention module is applied in the backbone of the UNet in the 
diffusion model  to maintain consistency among CKMs when inferring CKMs 
for multiple target BSs  simultaneously.
Specifically, disregarding the batch size and denoting the number of target BSs in 
set $I_{{\rm t}}$ as $|I_{{\rm t}}|$, 
it can be defined that the intermediate representation of $\mathbf{z}_t$ in the UNet is
$\tilde{\mathbf{X}}\in \mathbb{R}^{|I_{{\rm t}}|\times H_{\mathbf{X}} \times W_{\mathbf{X}} 
\times C_{\mathbf{X}} }$.
Through firstly flattening $\tilde{\mathbf{X}}$  by length, width, and  the number of  target BSs $|I_{{\rm t}}|$, we can get $\mathbf{X}\in \mathbb{R}^{N_{{\rm t}} 
\times C_{\mathbf{X}} }$, where $N_{{\rm t}} =|I_{{\rm t}}|\times H_{\mathbf{X}} \times W_{\mathbf{X}} $
denotes the length of tokens, while $C_{\mathbf{X}}$ denotes the dimension of tokens. 
Therefore, based on the principle of the attention mechanism, 
the  projection matrices can be obtained as \cite{vaswani2017attention,jaegle2021perceiver} 
\begin{equation}
  \mathbf{Q}^{\rm{s}}=\mathbf{X} \cdot \mathbf{W}^{\rm{s}}_Q, \mathbf{K}^{\rm{s}}=\mathbf{X} \cdot \mathbf{W}^{\rm{s}}_K,
  \mathbf{V}^{\rm{s}}=\mathbf{X} \cdot \mathbf{W}^{\rm{s}}_V,
\end{equation}
where $ \mathbf{W}^{\rm{s}}_Q\in  \mathbb{R}^{C_{\mathbf{X}}\times C_{\mathbf{X}}}$,
$ \mathbf{W}^{\rm{s}}_K\in  \mathbb{R}^{C_{\mathbf{X}}\times C_{\mathbf{X}}}$, and $ \mathbf{W}^{\rm{s}}_V\in  
\mathbb{R}^{C_{\mathbf{X}}\times C_{\mathbf{X}}}$ are learnable weight matrices.

The next step is to embed target BS locations  into $\mathbf{Q}^{\rm{s}}$ and $\mathbf{K}^{\rm{s}}$.
The corresponding rotation matrices are used for different target BSs. Specifically, for 
any target BS $i \in I_{{\rm t}}$, we have
\begin{equation}\label{eq:qttilde}
  \tilde{\mathbf{q}}_i=\begin{bmatrix}
   \mathbf{q}_1 \mathbf{R}_i^{T}  \\
    \vdots \\ 
    \mathbf{q}_{H_{\mathbf{X}}  W_{\mathbf{X}}} \mathbf{R}_i^{T} 
    \end{bmatrix} \in \mathbb{R}^{H_{\mathbf{X}}  W_{\mathbf{X}} \times C_{\mathbf{X}}},
\end{equation}
where $ \mathbf{R}_i$ is  the rotation matrix in (\ref{eq:Ri}), $\mathbf{q}_j \in 
\mathbb{R}^{1\times  C_{\mathbf{X}}}$ denotes the element from $ \mathbf{Q}^{\rm{s}}$.
Therefore, the query and key matrices after the embedding of  target BS locations
can be spliced as 
\begin{equation}\label{eq:Qtilde}
  \mathbf{\tilde{Q}}^{\rm{s}}=\begin{bmatrix}
    \tilde{\mathbf{q}}_1  \\
     \vdots \\ 
     \tilde{\mathbf{q}}_{|I_{{\rm t}}|}\end{bmatrix},
     \mathbf{\tilde{K}}^{\rm{s}}=\begin{bmatrix}
      \tilde{\mathbf{k}}_1  \\
       \vdots \\ 
       \tilde{\mathbf{k}}_{|I_{{\rm t}}|}\end{bmatrix},
\end{equation}
and the self-attention among  CKMs of multiple target BSs is defined as
\begin{equation}\label{eq:atten}
  {\rm Attention}(\mathbf{\tilde{Q}}^{\rm{s}},\mathbf{\tilde{K}}^{\rm{s}},\mathbf{V}^{\rm{s}})={\rm softmax}\big(  
    \frac{\mathbf{\tilde{Q}}\mathbf{\tilde{K}}^T}{\sqrt{C_{\mathbf{X}}}} \big) \cdot \mathbf{V}. 
\end{equation}

\subsubsection{Cross-Attention among Source and Target CKMs}
Cross-attention modules are added to the backbone of the UNet to 
maintain consistency between the  source CKMs and the target CKMs.
Similar to the analysis of self-attention, the batch size is disregarded, and the set of  source BSs is 
denoted as $I_{{\rm e}}$.
Intermediate representations of   CKMs of source BSs are considered as conditions for CKM inference across BSs, 
where the representations are obtained by a specific image encoder $\tau$ as
\begin{equation}
  \tilde{\mathbf{Y}}=\tau(\{ \mathbf{M}_i\}_{i\in I_{{\rm e}}} ) \in \mathbb{R}^{|I_{{\rm e}}|\times l \times C_{\mathbf{Y}}},
\end{equation}
where $l$ and $C_{\mathbf{Y}}$ denote  the length and dimension of tokens after $\tau$, respectively.
Then $\mathbf{Y} \in \mathbb{R}^{ N_{e} \times C_{\mathbf{Y}}}$ is obtained by further flattening $\tilde{\mathbf{Y}}$ by 
the number of  source BSs, where $N_{e}=|I_{{\rm e}}|\times l$.

To realize the cross-attention computation, the token dimension $C_{\mathbf{X}}$ of the intermediate representation $\mathbf{X}$ for 
the target CKMs needs to be   projected and aligned to that of the intermediate representations $\mathbf{Y}$ for the source CKMs.
Therefore, the corresponding learnable weight matrices are  derived as
\begin{equation}
  \mathbf{Q}^{\rm{c}}=\mathbf{X} \cdot \mathbf{W}^{\rm{c}}_Q, \mathbf{K}^{\rm{c}}=\mathbf{Y} \cdot \mathbf{W}^{\rm{c}}_K,
  \mathbf{V}^{\rm{c}}=\mathbf{Y} \cdot \mathbf{W}^{\rm{c}}_V,
\end{equation}
where $ \mathbf{W}^{\rm{c}}_Q\in  \mathbb{R}^{C_{\mathbf{X}}\times C_{\mathbf{Y}}}$,
$ \mathbf{W}^{\rm{c}}_K\in  \mathbb{R}^{C_{\mathbf{Y}}\times C_{\mathbf{Y}}}$, and $ \mathbf{W}^{\rm{c}}_V\in  
\mathbb{R}^{C_{\mathbf{Y}}\times C_{\mathbf{Y}}}$.

Furthermore, when performing BS location embedding, $\mathbf{\tilde{Q}}^{\rm{c}}$ is obtained from 
the rotation matrices  of  target BSs, while $\mathbf{\tilde{K}}^{\rm{c}}$ 
is obtained from 
the rotation matrices   of  source BSs.  Finally, the cross-attention 
among   CKMs of the target and source BSs is calculated by (\ref{eq:atten}).

Based on the detailed introduction to {\it BS-1-to-N}, Algorithm \ref{algorithm:training} and Algorithm \ref{algorithm:inference} demonstrate the training and inference processes, respectively.

\begin{algorithm}
  \caption{Training Process of {\it BS-1-to-N}}\label{algorithm:training}
  \KwIn{the target CKM $\mathbf{M}_{0}$ and  its location $\mathbf{l}_{0}$, the source  CKM  and location set $\{\mathbf{M}_{i},\mathbf{l}_{i} \}$}
  \KwOut{the   inferred CKM $\hat{\mathbf{M}}_{0}$, the trained UNet and image encoder $\tau$ }
  Transform $\mathbf{M}_{0}$ into the latent space as $\mathbf{z}_0$  through the VAE\;
  Transform $\mathbf{z}_0$ into $\mathbf{z}_T$ through the diffusion process\;
  Transform $\{\mathbf{M}_{i}\}$ into the intermediate representation $\mathbf{Y}$ with the image encoder $\tau$\;
  Construct the rotation matrices of $\mathbf{l}_{0}$ and $\{\mathbf{l}_{i} \}$\;
  \For{  \rm{each  denoising time step $t$ }   }
  {\For{  \rm{each  UNet layer }   }
    {
      Self-attention calculation with $\mathbf{\tilde{Q}}^{\rm{s}},\mathbf{\tilde{K}}^{\rm{s}},\mathbf{V}^{\rm{s}}$ after BSLE\; 
    Cross-attention calculation with $\mathbf{\tilde{Q}}^{\rm{c}},\mathbf{\tilde{K}}^{\rm{c}},\mathbf{V}^{\rm{c}}$ after BSLE\; }
    Calculate the denoising loss with (\ref{eq:loss}) and obtain $\mathbf{z}_{t-1}$\;
    Backward and update the parameters in UNet and $\tau$\;
    }
  Recover the $\hat{\mathbf{M}}_{0}$ from $\mathbf{z}_0$ through the VAE\;
  \end{algorithm}

\begin{algorithm}
    \caption{Inference Process of {\it BS-1-to-N}}\label{algorithm:inference}
    \KwIn{the target BS location set $\{\mathbf{l}_{0}\}$, the source  CKM  and location set $\{\mathbf{M}_{i},\mathbf{l}_{i} \}$}
    \KwOut{the   inferred CKM $\{\hat{\mathbf{M}}_{0}\}$ }
    Transform $\{\mathbf{M}_{i}\}$ into the intermediate representation $\mathbf{Y}$ with the image encoder $\tau$\;
    Construct the rotation matrices of $\{\mathbf{l}_{0}\}$ and $\{\mathbf{l}_{i} \}$\;
    \For{  \rm{each  denoising time step $t$ }   }
    {\For{  \rm{each  UNet layer }   }
      {
      Self-attention calculation with $\mathbf{\tilde{Q}}^{\rm{s}},\mathbf{\tilde{K}}^{\rm{s}},\mathbf{V}^{\rm{s}}$ after BSLE\; 
      Cross-attention calculation with $\mathbf{\tilde{Q}}^{\rm{c}},\mathbf{\tilde{K}}^{\rm{c}},\mathbf{V}^{\rm{c}}$ after BSLE\; }
      Calculate the denoising loss with (\ref{eq:loss}) and obtain $\mathbf{z}_{t-1}$\;
      }
    Recover the $\{\hat{\mathbf{M}}_{0}\}$ from $\mathbf{z}_0$ through the VAE\;
    \end{algorithm}

\section{Experiments Results}
In this section,  we present the training process of the proposed {\it BS-1-to-N} model  and validate its capability in the performance of cross-BS CKM inference. Furthermore, the application of cross-BS CKM inference in BS deployment optimization is demonstrated.

\subsection{Training Settings}

The  {\it BS-1-to-N} architecture is based on a specific implementation of the latent 
diffusion model, namely Stable Diffusion v1.5 \cite{rombach2022high}.
During the training process, Eschernet\cite{Kong2024eschernet}, 
targeting  the multi-view synthesis task, is used to generate  the starting point in order to reduce the training overhead.
Similarly, to improve the training efficiency, a lightweight image encoder $\tau$, 
called ConvNeXtv2-Tiny \cite{woo2023convnext}, is fine-tuned to encode the CKMs of  source BSs into 
corresponding  token features with smaller resolution.
These token features will be embedded with the corresponding BS locations, serving as conditions
used in the diffusion model to guide the CKM inference of  target BSs without physical environment maps.
The {\it BS-1-to-N} training is performed on RadioMapSeer \cite{DatasetPaper}, a radio map dataset 
based on OpenStreetMap obtained from cities such as London, Berlin, etc.
RadioMapSeer contains a total of 701 different physical environments, in which each 
physical environment is configured with 80 BSs at different locations, and a 
corresponding BS location map and channel gain map (CGM) are generated for each BS. 
CGMs of the dominant path model (DPM) are selected as  the training objects for 
CKM inference  across BSs, with the resolution of each CKM being $256\times256$.
In RadioMapSeer, the channel gain ranges from 47 dB to 147 dB. To map this range 
to the grayscale values of the CKM, a grayscale increment of 0.01 is defined to 
represent a 1 dB increase in channel gain.

During the training process, 90\% (630) of the physical environments with their 
corresponding CKMs are selected as the training set, while the remaining 10\% (71) of the 
physical environments  are selected as the validation set. The number $|I_{{\rm e}}|$  of source BSs
is set as 10, while the number  $|I_{{\rm t}}|$ of target BSs  is set as 5.
For each sample in the training process, we first select a specific physical environment. 
Then, 10 source BSs are selected from the 80 possible  BS locations in that physical environment, with 5 other BSs selected as target BSs. The locations of  source BSs are encoded and embedded with their CKMs as 
conditions, and the corresponding CKMs are inferred based on the locations of the 
target BSs. Note that the   physical environment map itself is not used as data 
for training. The learning rate employs a combination of linear warm-up and cosine annealing.
Some   main experiment parameters are shown in   Table \ref{table2p}.
\begin{table}[H]
  \caption{Main Experiment Parameters }
  \label{table2p}
  \centering
  \resizebox{0.7\columnwidth}{!}{
  \begin{tabular}{c|c}
  \hline
   Max training step    & 2000  \\
  Batch size&  10 \\
  $|I_{{\rm e}}|$  & 10   \\
  $|I_{{\rm t}}|$&  5    \\
  Learning rate range &$[1e^{-5},1e^{-4}]$\\
  Inference steps &50\\
  \hline
  \end{tabular}}
  \end{table}

The performance of cross-BS CKM inference based on {\it BS-1-to-N} can be characterized by 
different evaluation metrics. 
The performance of the trained {\it BS-1-to-N} is evaluated on the validation set. 
To evaluate the accuracy and reliability of the inferred CKMs 
from different perspectives, several typical metrics in the image domain are selected, 
including root mean squared error (RMSE), structural similarity index measurement (SSIM),
learned perceptual image patch similarity (LPIPS), and peak signal-to-noise ratio (PSNR).

\subsubsection{RMSE}
RMSE denotes the square root of the squared mean of the pixel differences between the 
inferred CKM and the real CKM. For the target BS 0, the RMSE between 
$\mathbf{M}_0$ and $\hat{\mathbf{M}}_0$ is 
\begin{equation}
  {\rm RMSE}=\frac{1}{L}\|\mathbf{M}_0-\hat{\mathbf{M}}_0\|_{F},
\end{equation}
where $\| \cdot \|_{F}$ denotes the  Frobenius norm.

\subsubsection{SSIM}
SSIM is a perceptual model that measures the degree of similarity between two images \cite{zhou2004image},
mainly considering   three key features, including luminance, contrast, and structure. 
Under the same RMSE, a higher SSIM value will be more in line with the intuitive 
perception of human eyes, where the value ranges from -1 to 1.
The computation of SSIM is based on the sliding window implementation, 
where
\begin{equation}
  {\rm SSIM}(a,b)=\frac{(2\mu_{a}\mu_{b}+c_1)(2\sigma_{ab}+c_2)}{(\mu_a^2+\mu_b^2+c_1)
  (\sigma_a^2+\sigma_b^2+c_2)},
\end{equation}
where $a$ and $b$ denote the same size windows of real CKM and inferred CKM, respectively.
$\mu_{a}$ and $\mu_{b}$ denote the mean of $a$ and $b$, while $\sigma_a^2$ and $\sigma_b^2$
denote the corresponding variance.
$c_1$ and $c_2$ are constants that maintain numerical stability.

\subsubsection{LPIPS}
LPIPS  is based on pre-trained deep learning networks to mine image features and 
evaluate the perceptual similarity between images to better simulate human visual perception \cite{zhang2018unreasonable}.
For CKM, the closer the value of LPIPS is to 0, the more similar the inferred CKM 
is to the real CKM.

\subsubsection{PSNR}
PSNR is an RMSE-based metric, whose core idea is to calculate the ratio of the maximum 
possible gain of the CKM to the   noise power.
On the basis of the calculated RMSE between the real CKM and the inferred CKM, 
the PSNR can be expressed as
\begin{equation}
  {\rm PSNR}=20 \cdot \rm \log_{10} (\frac{{\rm MAX}_{\mathbf{M}_0}}{{\rm RMSE}}),
\end{equation}
where ${\rm MAX}_{\mathbf{M}_0}$ denotes the theoretical maximum of CKM pixels.

\subsection{Cross-BS CKM Inference}
\subsubsection{Inference based on different number of source BSs}

\begin{figure*}[!t]
    \centering
      {\includegraphics[width=0.95\textwidth]{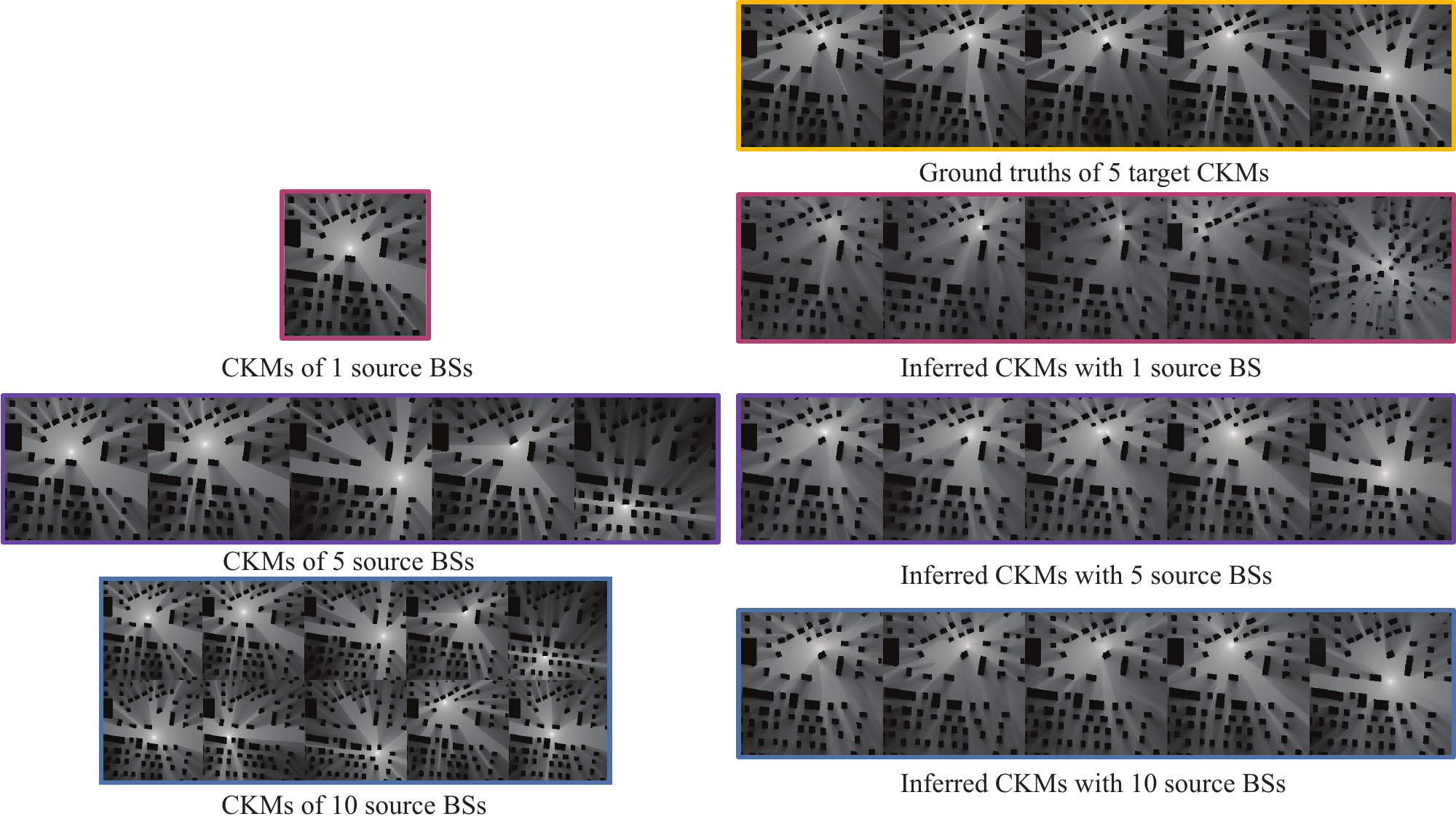}}%
    \caption{Comparison of the inferred CKMs of {\it BS-1-to-N} under different numbers of source BSs.}
    \label{fig:comparison1510}
  \end{figure*}

The cross-BS  inference is performed on the CKMs of  target BSs via pre-trained BS 
1-to-N using data from the validation set.  Note that the set of  target BSs is 
disjoint from the set of  source BSs.
The performance metrics of CKM inference for the same target BSs with different 
numbers of source BSs are shown in Table \ref{table_numbermetrices},
while a specific graphical presentation of the CKM
inference across BSs is shown in Fig. \ref{fig:comparison1510}. 
Since the physical environment map is unknown for {\it BS-1-to-N}, and we focus on the 
inference of channel knowledge, the building region is removed in the metric calculation.
From Fig. \ref{fig:comparison1510}, it can be seen that the mission of CKM inference 
across BSs can be well accomplished overall with the trained {\it BS-1-to-N}.
Without any physical environment information, {\it BS-1-to-N} enhances the awareness of the wireless environment and the corresponding channel knowledge by mining the CKMs of source BSs. As shown in the graphical results, the inferred CKMs clearly demonstrate the unknown physical properties, along with the attenuation and blocking properties of the channel gain.
Meanwhile, by comparing the scenarios with different numbers of source BSs, it can be found that as the number of source BSs and their corresponding CKMs increase, the inferred CKMs of  target BSs become more accurate.
This is reasonable since each additional source CKM implies more channel knowledge, from which {\it BS-1-to-N} can further learn more about the environment prior for further cross-BS CKM inference.
As the number of source BSs increases, the marginal gain of improving cross-BS CKM inference performance also decreases, which is evident from the comparison between the scenarios with 5 source BSs and 10 source BSs.

\begin{table}[H]
  \caption{Evaluation Metrics with Different source BS Numbers }
  \label{table_numbermetrices}
  \centering
  \resizebox{0.85\columnwidth}{!}{
  \begin{tabular}{c|cccc}
  \hline
      & RMSE$\downarrow$ &SSIM$\uparrow$ & PSNR$\uparrow$ &LPIPS$\downarrow$  \\
  \hline
  1 source BSs& 0.1056 & 0.8549 &19.8783&0.2247 \\
   5 source BSs &  0.0649 & 0.8980 & 24.0652&0.1342  \\
  10 source BSs& 0.0615 &0.9025&25.5542 &0.1278 \\
  \hline
  \end{tabular}}
  \end{table}

\subsubsection{Inferred CKM comparison   between {\it BS-1-to-N} and other benchmarks}

To further evaluate the CKM inference performance  for {\it BS-1-to-N}, 
the inferred CKMs are compared with those of other benchmark schemes.
The following typical schemes are selected 
as benchmarks.
\begin{itemize}
  \item UNet-based CKM inference scheme across BSs, where the weighted sum of CGM information and  source BS location information  
  with different modalities is adopted as features, expressed  as \cite{dai2024generating}
  
  \begin{equation}
    \mathbf{M}_{i}=(1-\omega)\mathbf{G}_i +\omega \mathbf{M}_{\mathrm{BS},i},
  \end{equation}
  where $\mathbf{G}_i$ and $\mathbf{M}_{\mathrm{BS},i}$ are the CGM and the  location map
  of BS $i$. 

  \item Cross-BS CKM construction scheme with distance-based weighting, 
  where the CKM of the target BS is calculated as the weighted sum of the source BS CKMs, expressed as
  
  \begin{equation}
    \hat{\mathbf{M}}_{0}=\sum_{i=1}^{I} w_{i}\mathbf{M}_i
    =\sum_{i=1}^{I} \frac{e^{-\gamma d_{i}}}{\sum_{j=1}^{I} e^{-\gamma d_{j}} }\mathbf{M}_i,
  \end{equation}
where $d_{i}$ denotes the Euclidean distance between BS $i$ and the target BS 0.
The weights $\{  w_{i} \}$ are normalized according to the distance set $\{ d_{i}\}$
with the weight parameter $\gamma=0.1$.
\end{itemize}

\begin{figure*}[!t]
    \centering
      {\includegraphics[width=0.95\textwidth]{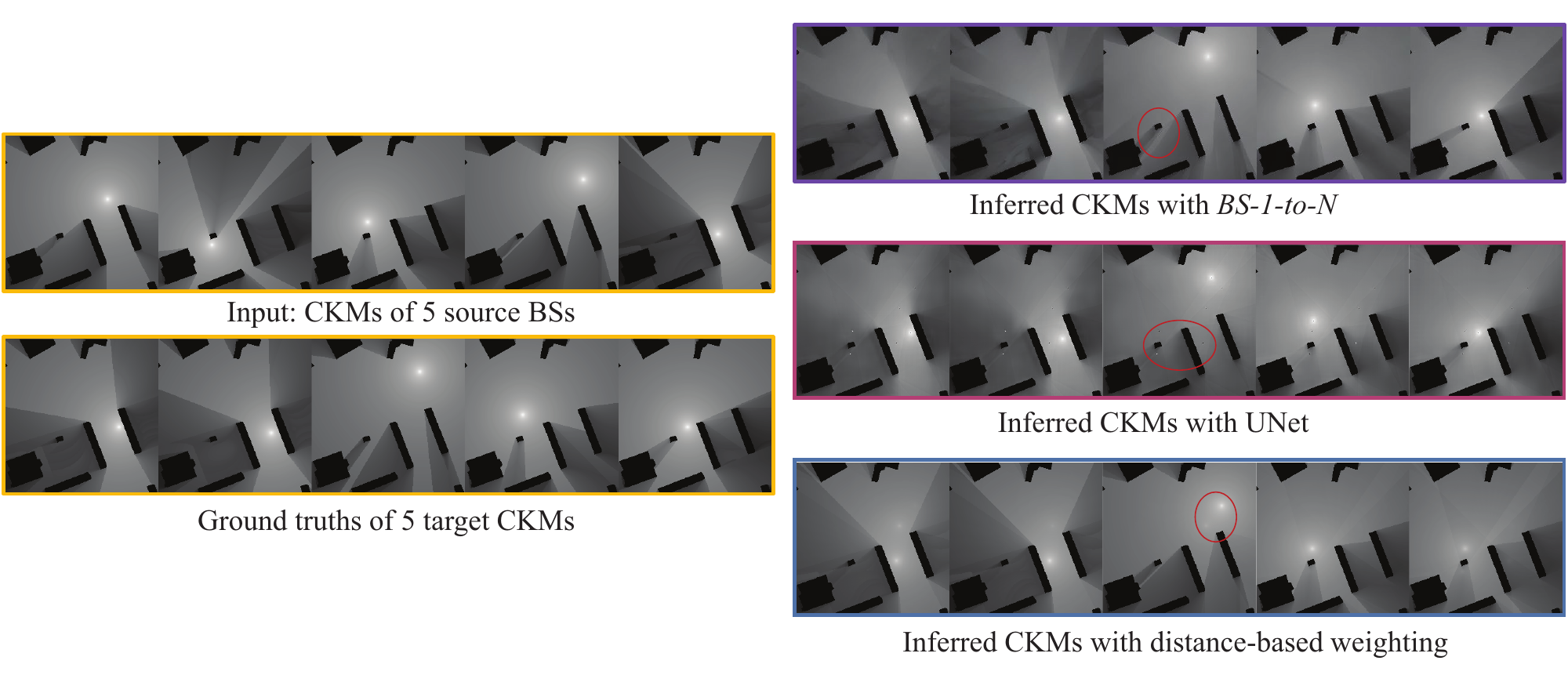}}%
    \caption{Comparison of the inferred CKMs of {\it BS-1-to-N} with other benchmarks under 5 source BSs.}
    \label{fig:comparison5_bench}
\end{figure*}

The graphical presentation and performance metrics of CKM inference across BSs are 
shown in Fig. \ref{fig:comparison5_bench} and Table \ref{table_benchmetrices}, respectively.
Notice that both {\it BS-1-to-N} and the other benchmarks pick the  same 5 source BSs and 
infer the CKMs of the   same 5 target BSs.
Although {\it BS-1-to-N} does not dominate the evaluation metrics with the other 
benchmarks, a different conclusion can be observed in the CKM presentation in Fig. \ref{fig:comparison5_bench}.
For the distance-based weighting scheme, although inferred CKMs present the blocking 
property of channel gain under the influence of buildings, 
the scheme cannot accurately find the location of the target BS and essentially 
reuses the CKM of  source BSs.
As a result, the distance-based weighting scheme performs particularly poorly with a limited number 
of source BSs, and can even result in fake BSs as circled by the red line  in Fig. \ref{fig:comparison5_bench}.
For the UNet-based scheme, the BS location information and CKM information of different
modalities are directly weighted; hence, it is difficult to distinguish their differences 
during the learning process, which makes it difficult to learn the wireless environment.
Therefore, when the number of source CKMs is limited,
the inferred CKMs with the UNet-based scheme are smoother on the one hand, which makes it difficult to show the blocking property of channel gain, and on the other hand, noise points may also occur as  circled by the red circle in Fig. \ref{fig:comparison5_bench}.

For {\it BS-1-to-N}, despite the different modalities, the location information of  BSs is 
embedded into the corresponding CKMs via BSLE and further guides the CKM inference 
across BSs via the attention mechanism.
Compared with the benchmarks, the trained {\it BS-1-to-N} has a significantly better awareness 
of the wireless environment and is able to accurately capture the target BS location and 
the
wireless propagation characteristics under the influence of the physical environment.
Meanwhile, it is important to note that despite the differences in evaluation metrics 
between {\it BS-1-to-N} and the UNet-based scheme, 
UNet was trained with CKMs of 79 source BSs as inputs, while {\it BS-1-to-N} used only 
10 source BSs during the training.
If the number of source BSs used for training can be further increased 
with hardware support, the performance of {\it BS-1-to-N} will be improved accordingly.

\begin{table}[H]
  \caption{Comparison of the inferred CKMs with Other Benchmarks  Under 5 source BSs   }
  \label{table_benchmetrices}
  \centering
  \resizebox{0.85\columnwidth}{!}{
  \begin{tabular}{c|cccc}
  \hline
      & RMSE$\downarrow$  &SSIM$\uparrow$ & PSNR$\uparrow$ &LPIPS$\downarrow$  \\
  \hline
  {\it BS-1-to-N}&  0.0649 & 0.8980 & 24.0652&0.1342  \\
   UNet  &  0.0354 & 0.9449 & 29.8511&0.0672  \\
  Distance-based weighted& 0.0785 &0.9056&24.6489 &0.0576 \\
  \hline
  \end{tabular}}
  \end{table}

\begin{figure}[!t]
    \centering
      {\includegraphics[width=0.9\columnwidth]{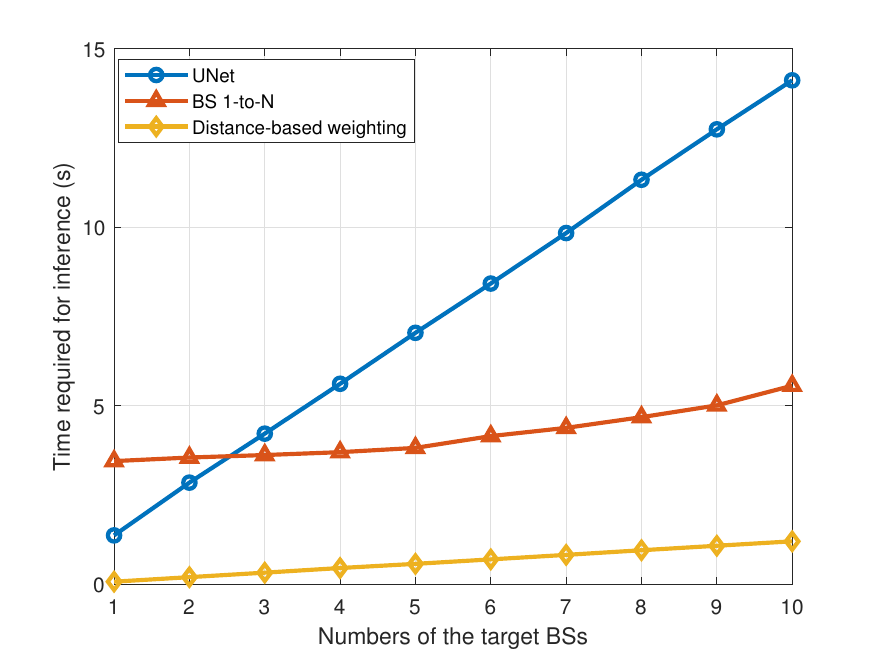}}%
    \caption{Comparison of the inference time under different numbers of target BSs.}
    \label{fig:comparison_time}
\end{figure}

Fig. \ref{fig:comparison_time} shows the execution time for cross-BS CKM 
inference under different numbers of target BSs.
The number of source BSs is set as 5, and the CKM inference is performed on a single NVIDIA Tesla V100 (16 GB).
The distance-based weighting scheme clearly requires the shortest inference time 
due to its simple computation process.
The inference time cost by the UNet-based scheme is first lower than 
and rapidly exceeds that of {\it BS-1-to-N} as the number of target BSs increases.
This is explained by the fact that since the UNet architecture is fixed, 
CKM inference is separately executed for each target BS, making the total inference
time proportional to the number of target BSs.
In contrast, the improvement of the attention module enables {\it BS-1-to-N} to learn the 
interactions between any number of the source CKMs and the target CKMs,
ultimately realizing CKM inference from BS 1 to BS $N$.
This architecture design eliminates the need for {\it BS-1-to-N} to perform repetitive 
inference on CKMs of multiple target BSs and maintains a similar inference time 
growth rate as the distance-based weighting scheme.

\subsection{BS Deployment Optimization based on CKM Inference}

To validate the application prospects, CKM inference across BSs is considered to 
be applied to the deployment optimization of potential new BSs.
We select  a specific physical environment map as an example and optimize the 
location of a potential new BS by inferring the corresponding CKM based on {\it BS-1-to-N}.



\begin{figure}[!t]
  \centering
    {\includegraphics[width=0.9\columnwidth]{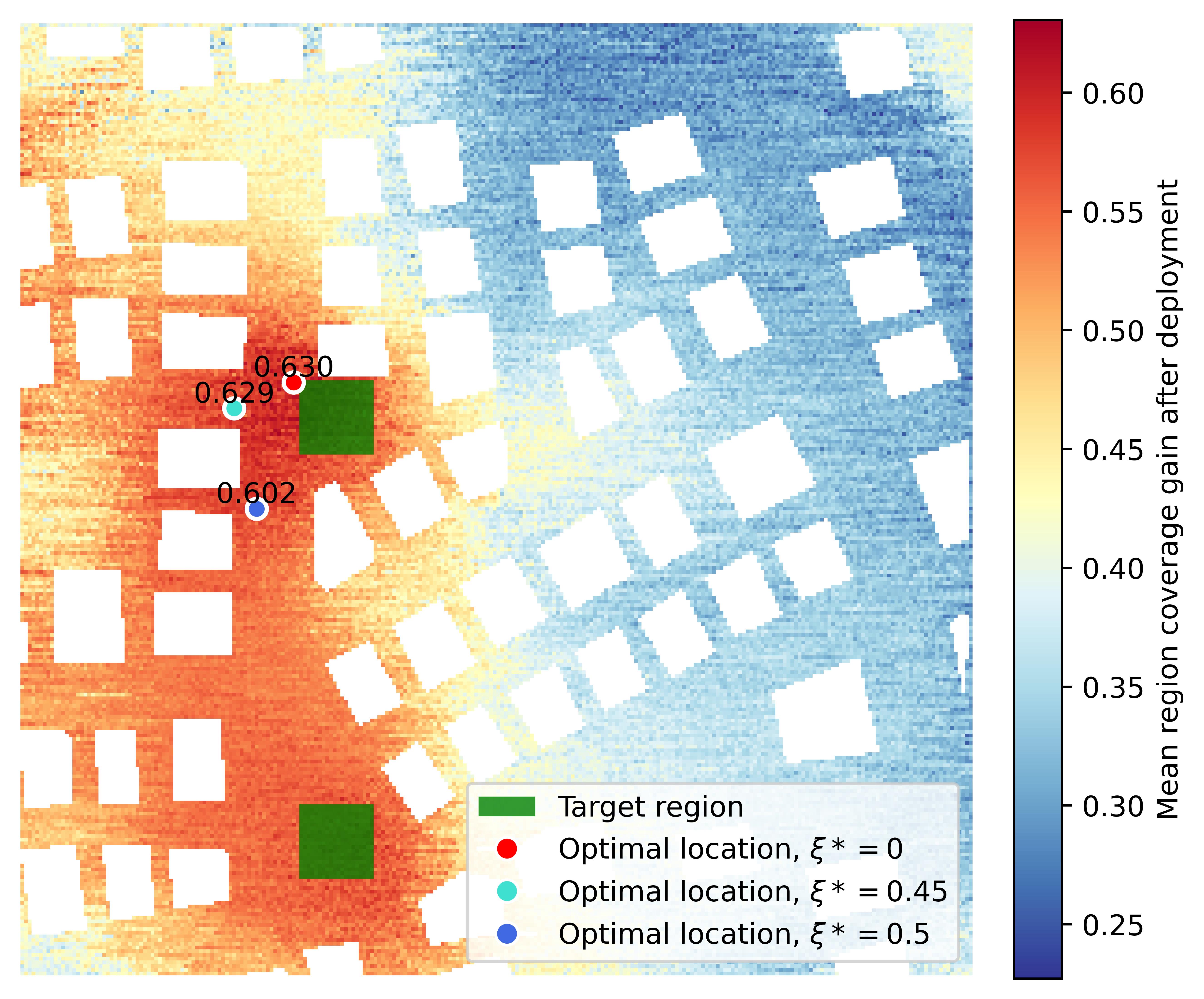}}%
  \caption{Multi-target region coverage with different target BS locations.}
  \label{fig:Multi-target}
\end{figure}

As shown in Fig. \ref{fig:Multi-target}, consider two target regions represented by green. 
The goal of coverage optimization is to find the appropriate deployment location of 
the target BS 
that can guarantee good coverage for the target regions.
Specifically, 10 fixed source BSs are randomly selected as input, and the 
trained {\it BS-1-to-N} traverses all potential target BS locations to infer the 
corresponding CKMs and evaluate the coverage performance for  target regions.
During the deployment optimization, for each potential location of the target BS, 
the minimum channel gain $\xi_{\rm{min},1}$ and $\xi_{\rm{min},2}$  of the
target regions  1 and 2 are stored separately, 
along with the mean channel gain $\bar{\xi}$  for both target regions.
In  Fig. \ref{fig:Multi-target}, the value of each point represents the mean channel 
gain from the target BS at that location
to  the target regions. 
The channel gain values have been normalized, and the white patch denotes buildings in the map.
Effective BS deployment requires good coverage performance of target regions.
Therefore, given a channel gain threshold $\xi*$, the potential locations of the target 
BS need to satisfy $\xi_{\rm{min},1}\geq \xi*$ and $\xi_{\rm{min},2}\geq \xi*$.
Further, after meeting the minimum coverage threshold requirement,
the optimal deployment location of the target BS that maximizes the mean channel gain 
of multiple target regions is found.
The optimal location of the target BS under different channel gain thresholds is indicated by blue and orange dots, respectively.
As the threshold  $\xi*$ increases, the optimal location of the target BS seeks a 
compromise between the coverage performance of different target regions based on the 
inferred CKMs.
Therefore, based on the inferred CKMs from {\it BS-1-to-N}, the deployment location of potential 
target BSs can be optimized quickly and efficiently.


\section{Conclusion}
This paper proposed  a generative environment-aware cross-BS CKM inference framework 
called {\it BS-1-to-N}.  
Cross-BS CKM inference was modeled as a conditional generation problem, while 
BSLE and attention mechanisms were used to capture the implicit correlation of 
CKMs between different BSs. Therefore,
{\it BS-1-to-N} achieved flexible   CKM inference across an  arbitrary number of BSs 
without the  reconstruction of the physical or wireless environment.
Experiment results showed that {\it BS-1-to-N} performs accurate CKM inference with a 
limited number of source BSs and can efficiently capture wireless environment 
characteristics.
The optimization application of BS deployment further verified the  value of cross-BS CKM inference 
with {\it BS-1-to-N} in dense networks for environment-aware communication.
In the future, our work will focus on enhancing the robustness of cross-BS CKM inference 
in dynamic environments.
\begin{appendices}

\end{appendices}


\bibliographystyle{IEEEtran}

\bibliography{IEEEabrv,ref}


\end{document}